# Highly Amplified Light Transmission in Parity-Time Symmetric Multilayered Structure

## JYOTI PRASAD DEKA[1] and AMARENDRA K. SARMA[1,*]

[1]Department of Physics, Indian Institute of Technology Guwahati, Guwahati - 781039, Assam, India
*Corresponding author: aksarma@iitg.ernet.in

**We propose a parity-time symmetric dielectric-nanofilm-dielectric multilayered structure that could facilitate highly amplified transmission of optical power in the infrared spectrum. We have theoretically studied our model using the transfer matrix formalism. The reflection and the transmission coefficients of the S-matrix are evaluated. The theoretical results are validated by FDTD numerical simulation. We have shown how the thickness of the layers and the gain/loss coefficient of the active layers could generate spectral singularities in the S-matrix and how these singularities could be exploited to achieve amplified transmission of a single wavelength through the structure.**

## 1. INTRODUCTION

The concept of Parity-Time (PT) symmetry, which originally belongs to quantum mechanics, has recently attracted tremendous attention, in the fields of optics and photonics owing to many possible unique photonic device functionalities and applications. In quantum mechanics, hermiticity of Hamiltonians plays a very important role since only Hermitian Hamiltonians can exhibit real eigen-spectra and as such, they deal with physical observables. However, Bender and Boettcher's seminal work [1] in the year 1998 showed that a class of non-Hermitian Hamiltonians can exhibit entirely real eigen spectra, provided they are invariant under the operation of the PT (Parity and Time Reversal) operator. The distinctive property in such Hamiltonians is that they can undergo spontaneous symmetry breaking and consequently, their eigen-spectra transform from entirely real to complex. This refers to as the so-called 'phase transition' from the exact to the broken-PT phase. These symmetry breaking points (also known as exceptional points) are termed as the PT threshold of the Hamiltonian.

The idea that optics could provide the platform for the experimental realization of PT symmetric potentials was first proposed by Chirstodoulide's group [2]. They proposed that evanescently coupled dielectric optical slab based waveguides with balanced gain and loss can offer a feasible way to experimentally validate these mathematical theories. This proposition was conceivable due to the isomorphism between the paraxial equation of diffraction and the Schrodinger's wave equation. The refractive index potential $n(\vec{r})$ in the former and the potential function $V(x)$ in the later serve as the connection for this shared isomorphism. In addition to this, the refractive index potential must satisfy the condition $n(\vec{r}) = n^*(-\vec{r})$. This means that the real part of $n(\vec{r})$ must be an even function of $\vec{r}$ while the imaginary part an odd function of $\vec{r}$. The first set of experimental works was reported soon [3, 4] resulting in a plethora of research activities in PT-symmetric optics. Ruter *et al.* [3] studied a coupled optical waveguide configuration on LiNbO$_3$ substrate. One of the channel was optically pumped to provide gain and the other channel exhibits an equal amount of loss. They reported that the gain/loss coefficient of the waveguides serves as the measure for exact or broken PT symmetry in their configuration. Since then, numerous phenomena have been reported in the field of parity time symmetric optics. To cite, some of them include: onset of chaos in optomechanical systems [5], optical mesh lattices [6-8], modulational invisibility in complex media [9], Peregrine soliton dynamics [10], plasmon excitation [11, 12], unidirectional reflectionlessness, invisibility and non-reciprocity in periodic structures [13-17], whispering gallery modes [18], microring resonators [19-21], optical oligomers [22-25] and so on.

Unidirectional reflectionlessness, invisibility and non-reciprocity have been studied in several works till now. In [15], unidirectional invisibility in a PT symmetric Braggs grating structure was reported. Using the S-matrix formalism, it is found that optical power, when launched from either of the two ports, showed entirely different behavior in the reflection and transmission properties of the structure. More recently, the same formalism has been used to study unidirectional reflectionlessness and invisibility [14] in a multilayered structure. In our work, we propose a multilayered structure that can serve as a single wavelength filter with high amplification in the exact PT symmetric state. The novelty of the work lies in the simplicity of the structure, which could possibly be realized in experiments fairly easily. We have worked out the entire theoretical framework from the first principles and validated the theoretical results with FDTD simulation.

The article is organized as follows. In section II, starting with the boundary conditions of the incident and reflected fields at the interfaces, the transfer matrix for the proposed structure is worked out. In section III, the eigen-spectra of the transfer matrix is numerically evaluated followed by a study on the transmission and the reflection characteristics of the structure numerically as well as by using FDTD simulation. After that, it is shown how modifying the length of the layers and its gain/loss coefficient could impact the transmission of the radiation at 1550 nm, the so-called telecommunication wavelength. And finally in section IV, we have concluded our work.

## 2. THEORETICAL MODEL

Parity time symmetry was first studied experimentally in a coupled optical waveguide system known as the dimer [4]. The optical dimer is, in fact, the simplest configuration in a much larger class of coupled optical waveguide systems. In such systems, either the Hamiltonian or the Transfer matrix formalism can describe the evolution of optical power in the linear non-dispersive regime. In the former, the equations [4] governing the evolution of optical power are given by

$$i\frac{d}{dz}\begin{pmatrix}\psi_1\\\psi_2\end{pmatrix} = H_{PT}\begin{pmatrix}\psi_1\\\psi_2\end{pmatrix} \quad (1)$$

Here, $H_{PT} = \begin{pmatrix} i\gamma & C \\ C & -i\gamma \end{pmatrix}$ and its eigenvalues are given by $\lambda_H = \pm\sqrt{C^2 - \gamma^2}$, where $\gamma$ is the loss/gain parameter of the optical fibers and $C$ is the coupling parameter. The system is in the unbroken PT-symmetric regime when $\gamma < C$ and in the broken regime when $\gamma > C$. The critical point of symmetry breaking, also known as the PT threshold, is located at $\gamma_{PT} = C$.

In the transfer matrix formalism [26], these equations are needed to be solved first analytically using the initial conditions of Eq. 1. This is followed by the construction of the transfer matrix describing the evolution of optical power. Suppose, we launch electric field amplitudes $\psi_1(0)$ and $\psi_2(0)$ at the input ports of the waveguides. Then after propagating a distance $z = L_c$, the field amplitudes in the unbroken regime will be given by

$$\begin{pmatrix}\psi_1(L_C)\\\psi_2(L_C)\end{pmatrix} = T\begin{pmatrix}\psi_1(0)\\\psi_2(0)\end{pmatrix} \quad , \quad (2)$$

where $T$ is the transfer matrix of the configuration. The matrix elements are given by: $T_{11} = \cos(pL_C) + g/p\sin(pL_C)$, $T_{12} = T_{21} = -iC/p\sin(pL_C)$ and $T_{22} = \cos(pL_C) - g/p\sin(pL_C)$ and $p = -\sqrt{C^2 - \gamma^2}$. The parameter $L_C$, termed as the coupling length, is given by, $L_C = \pi/2C$. Similarly, in the broken symmetry regime, we have $p = \sqrt{\gamma^2 - C^2}$ and all the sinusoidal terms in the transfer matrix will be replaced by their hyperbolic counterparts. Let us denote the eigenvalues of the transfer matrix as $\lambda_T$.

Fig. 1 depicts the eigen-spectra for both the approaches. The PT threshold of the system, regardless of whether we use the Hamiltonian or the Transfer matrix formalism, remains unchanged, i.e. at $\gamma_{PT} = C$. The two approaches can only be distinguished in their eigen-spectra. In the Hamiltonian formalism, the eigen-spectra is entirely real in the unbroken regime and purely imaginary in the broken symmetry regime.

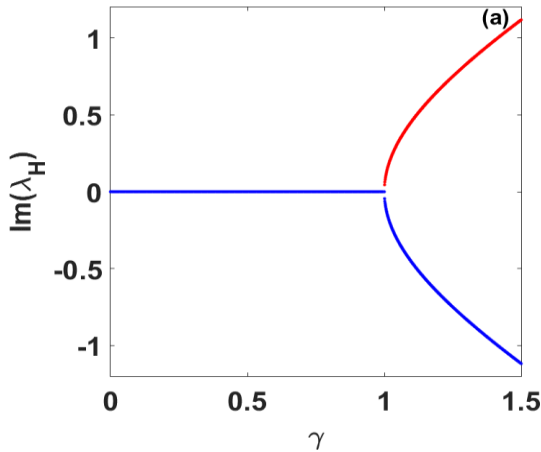

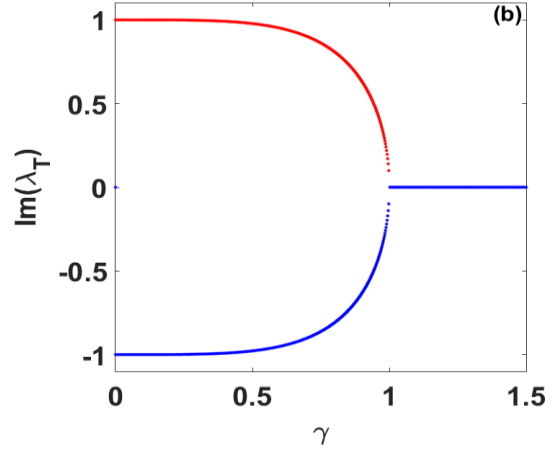

**Fig. 1.** Eigen spectra of the (a) Hamiltonian $H_{PT}$ and (b) Transfer Matrix $T$ for $C = 1$.

On the other hand, in the transfer matrix formalism, the eigen-spectra are complex in the unbroken regime and real in the broken regime. To clarify this issue further, in Fig. 2(a) and (b), we plot the magnitude of the eigenvalues and the real part of the complex eigenvalues as a function of the loss/gain parameter.

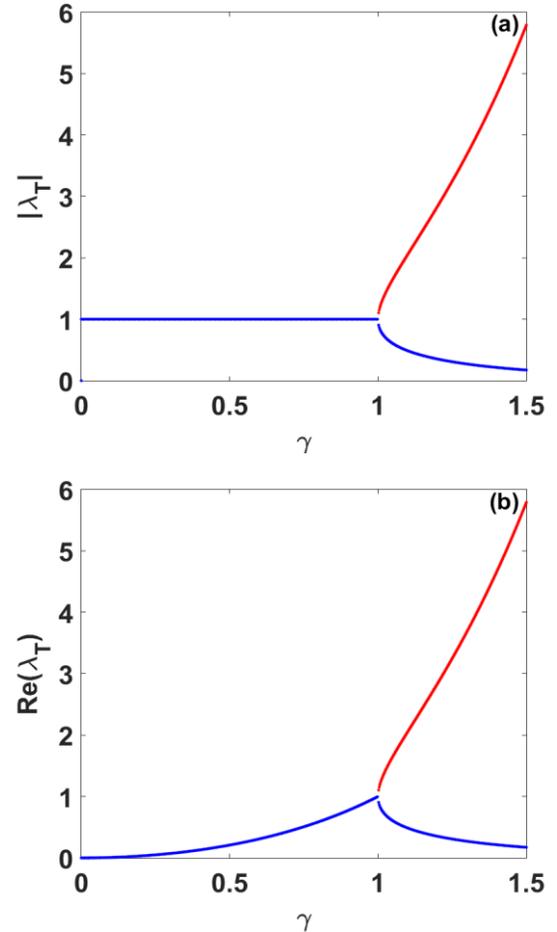

**Fig. 2.** (a) $|\lambda_T|$ (b) $\text{Re}(\lambda_T)$, as a function of loss/gain parameter

It could be seen from Fig. 2(a) that the magnitude of the eigenvalues are equal in the unbroken regime and beyond the PT-threshold, they are no longer equal. This happens owing to the bifurcation of the real component of the eigenvalues at the PT-threshold as could be seen in Fig. 2(b).

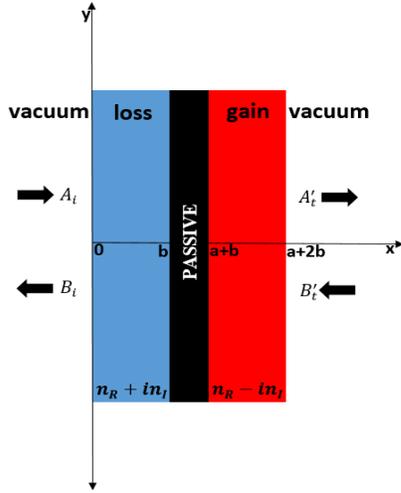

**Fig. 3.** Schematic of the dielectric-nano-film-dielectric multilayered structure. '$b$' is the length of the active regions (colored red and blue) and '$a$' is the length of the nano-film region (colored black).

In this work, we have considered a dielectric-nano-film-dielectric multilayered (DND) structure (Fig. 3). The dielectric layers are active media while the nano-film is a passive medium. The length of the nano-film is very small compared to the loss and the gain region. The structure is surrounded by vacuum on all sides. Considering normal incidence of the incoming fields and the x-axis as its propagation direction, boundary conditions at the input port, results:

$$A_i + B_i = A'_1 + B'_1 \qquad (3)$$
$$A_i - B_i = n_L(A'_1 - B'_1) \qquad (4)$$

Here, $A_i$ and $B_i$ are the forward and backward propagating wave amplitudes to the left of the input port while $A'_1$ and $B'_1$ are the same to the right of the input port. $n_L$ is the refractive index of the loss-layer and is given by $n_L = n_R + in_I$, where $n_R$ and $n_I$ are the real and the imaginary component of the refractive index respectively. Applying boundary conditions, in a similar way, for other interfaces we can express the forward and backward wave amplitudes at the output port of the structure in terms of its counterpart at the input port as follows [26-29]:

$$\begin{pmatrix} A'_t \\ B'_t \end{pmatrix} = T \begin{pmatrix} A_i \\ B_i \end{pmatrix} \qquad (5)$$

where, $T = D_i^{-1}D_3P_3D_3^{-1}D_2P_2D_2^{-1}D_1P_1D_1^{-1}D_i$ is the transfer matrix of the structure. Here, $D_i = \begin{pmatrix} 1 & 1 \\ 1 & -1 \end{pmatrix}$, $D_1 = \begin{pmatrix} 1 & 1 \\ n_L & -n_L \end{pmatrix}$, $D_2 = \begin{pmatrix} 1 & 1 \\ n_{film} & -n_{film} \end{pmatrix}$ and $D_3 = \begin{pmatrix} 1 & 1 \\ n_G & -n_G \end{pmatrix}$. $n_{film}$ and $n_G$ are the refractive indices for the film and the gain-layer respectively, with $n_{film} = n_{R,film} + in_{I,film}$ and $n_G = n_R - in_I$. On the other hand, $P_1 = \begin{pmatrix} e^{ik_1b} & 0 \\ 0 & e^{-ik_1b} \end{pmatrix}$, $P_2 = \begin{pmatrix} e^{ik_2a} & 0 \\ 0 & e^{-ik_2a} \end{pmatrix}$ and $P_3 = \begin{pmatrix} e^{ik_3b} & 0 \\ 0 & e^{-ik_3b} \end{pmatrix}$ are the propagation matrices of loss, film and gain layer respectively, while $k_1 = n_L\omega/c$, $k_2 = n_{film}\omega/\square$ and $k_3 = n_G\omega/c$ are the corresponding wave numbers. Finally, the transfer matrix could be expressed as:

$$T = T_3T_2T_1T_0/16 \qquad (6)$$

where
$$T_3 = \begin{pmatrix} 1 + n_G & 1 - n_G \\ 1 - n_G & 1 + n_G \end{pmatrix},$$
$$T_2 = \begin{pmatrix} e^{ik_3b}(1 + n/n_G) & e^{ik_3b}(1 - n/n_G) \\ e^{-ik_3b}(1 - n/n_G) & e^{-ik_3b}(1 + n/n_G) \end{pmatrix},$$
$$T_1 = \begin{pmatrix} e^{ik_2a}(1 + n_L/n) & e^{ik_2a}(1 - n_L/n) \\ e^{-ik_2a}(1 - n_L/n) & e^{-ik_2a}(1 + n_L/n) \end{pmatrix}$$

and
$$T_0 = \begin{pmatrix} e^{ik_1b}(1 + 1/n_L) & e^{ik_1b}(1 - 1/n_L) \\ e^{-ik_1b}(1 - 1/n_L) & e^{-ik_1b}(1 + 1/n_L) \end{pmatrix}.$$

## 3. RESULTS AND DISCUSSION

Analytical calculations for the eigenvalues and their corresponding eigenvectors of the transfer matrix $T$ are a rather tedious job. That is why we have numerically evaluated the eigen-spectra of the transfer matrix. We consider the dielectric of the structure to be made of silica ($n_R = 1.4657$) while the nano-film to be made of sapphire ($n_{film} = 1.7462$). The sapphire nano-film is considered to be a passive medium. This means $n_{I,film} = 0$. The dielectrics on the two sides of the nano-film are doped to have equal amount of loss and gain respectively. The length, $b$, of the silica layers is assumed to be 2.49 μm and that of the nano-film is, $a = 0.02$ μm. Fig. 4 depicts the eigen-spectra of the structure when it is excited from (a) loss-port and (b) gain-port, with a radiation centered at the wavelength 1550 nm.

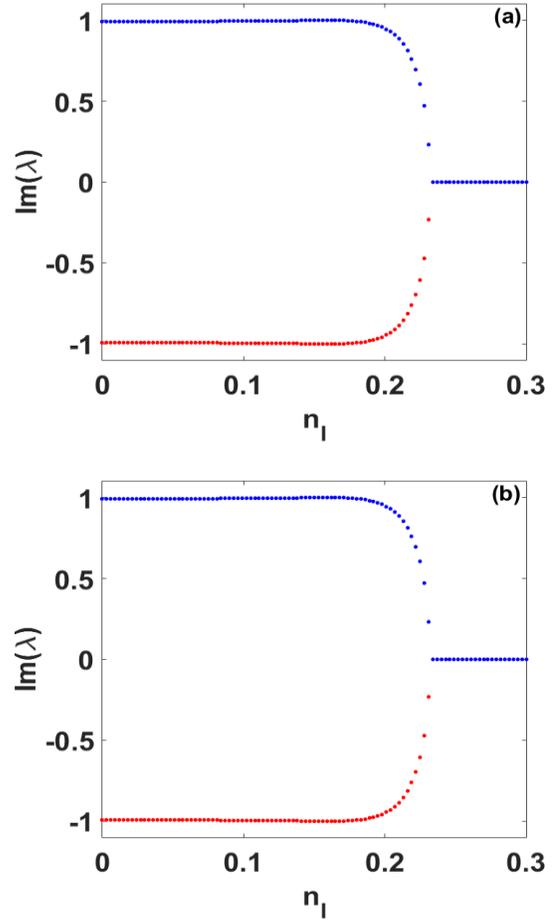

**Fig. 4**. Eigen-spectra of the transfer matrix for excitation of (a) loss port and (b) gain port.

From Fig. 4, we observe that as $n_I$ is increased, the imaginary component of the transfer matrix eigenvalues ceases to be imaginary beyond the PT-threshold. Moreover, the eigen-spectra do not depend on whether optical power is launched from the loss port or the gain port. For our chosen system parameters, the PT threshold is found to be at $n_I \approx 0.23$.

The transmission and reflection coefficients of the S-matrix could be obtained from the elements of the transfer matrix $T_{ij}$. The transmission coefficient is given by $T' = \left|\frac{1}{T_{22}}\right|^2$ and the reflection coefficient by $R' = \left|-\frac{T_{21}}{T_{22}}\right|^2$. The normalized transmission and the reflection coefficients as a function of the loss/gain parameters, for both loss and gain port excitation, are plotted in Fig. 5 and 6 respectively. The transmission characteristics of the DND structure, as depicted in Fig.5, are found to be independent of the launch conditions. There is near complete transmission when one operates below the PT-threshold, while above the PT threshold, the transmission coefficient decreases exponentially to zero. On the other hand, as shown in Fig. 6, the reflection characteristics exhibit significantly different behavior.

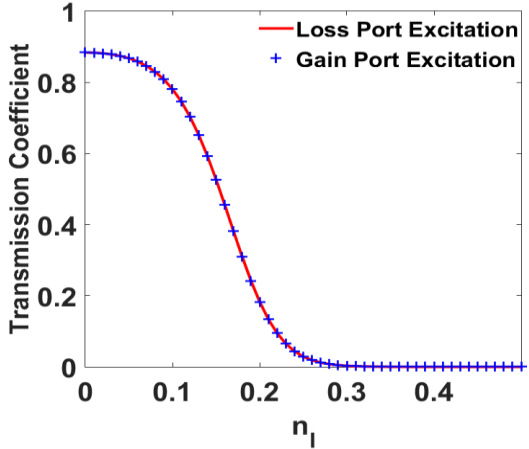

**Fig. 5.** Transmission Coefficient ($T'$) v/s Gain/Loss Coefficient $n_I$ for excitation of loss or gain port.

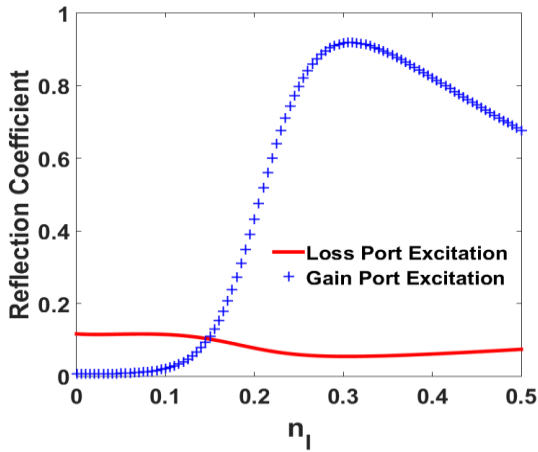

**Fig. 6.** Reflection Coefficient ($R'$) v/s Gain/Loss Coefficient $n_I$ for excitation of loss or gain port.

If operated above the PT-threshold, and the radiation is launched from the gain port, the reflection coefficient increases to very high values, implying significant reflection. This is supported by very low value of the transmission coefficient as well.

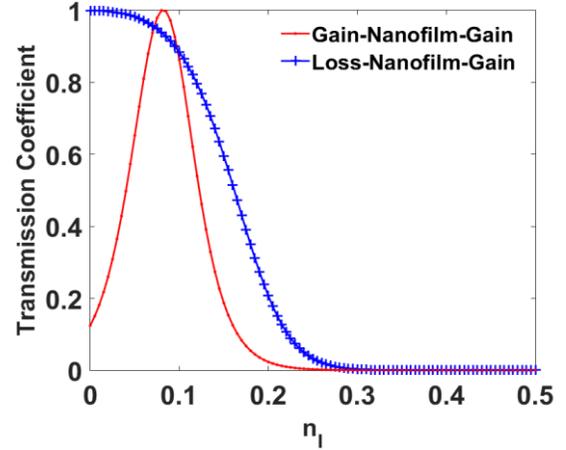

**Fig. 7.** Transmission Coefficient ($T'$) v/s Gain/Loss Coefficient $n_I$.

In Fig. 7, we have shown how our structure is superior in comparison to a gain-nanofilm-gain structure. We can see that for a gain-nano-film-gain structure, the transmission coefficient rises sharply initially as we increase $n_I$. But then it decreases to zero, whereas the transmission coefficient of our structure remains high in the unbroken PT regime. Only above the PT threshold, it decays to zero.

FDTD simulations validate these theoretical results. Fig. 8 and 9 depicts, for below and above the PT-threshold regime respectively, the contour plot for time-averaged power propagation in the DND structure. $P_I$, $P_R$ and $P_O$ are, respectively, the input, the reflected and the output optical power. In order to understand these plots, it is important to note the followings about the color bar. The positive (negative) values in the color bar signify that the optical power is propagating in the positive (negative) x-direction. The plots clearly illustrate the dependence on initial launch condition of the optical radiation and the PT-threshold. If one operates below the PT threshold, as shown in Fig. 8, near complete transmission of optical radiation could be observed. However, the nature of evolution of optical power is different for both the launch conditions. The structure, as depicted in Fig. 9, exhibits completely different behavior if one operates above the PT-threshold. It is possible to achieve near complete transmission when the radiation is launched from the loss port while almost complete reflection of optical radiation is obtained when the structure is excited from the gain port. Thus, the DND structure exhibits unidirectional invisibility if operated above the PT-threshold.

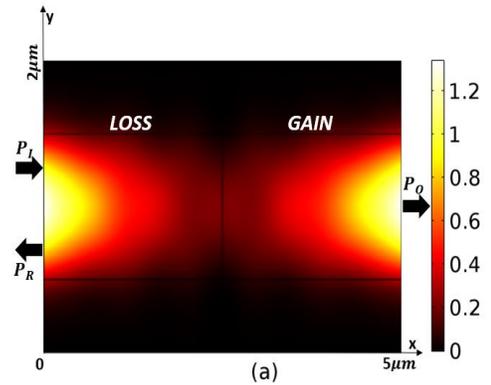

(a)

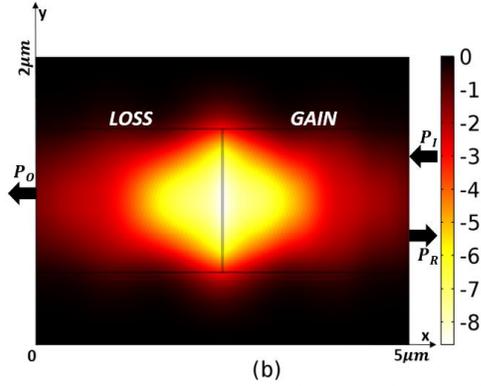

**Fig. 8.** Contour plot of 'Time-averaged power flow (in the unit of $10^6 W/m^2$)' for initial excitation of (a) loss port and (b) gain port when $n_I = 0.1$ (below PT threshold).

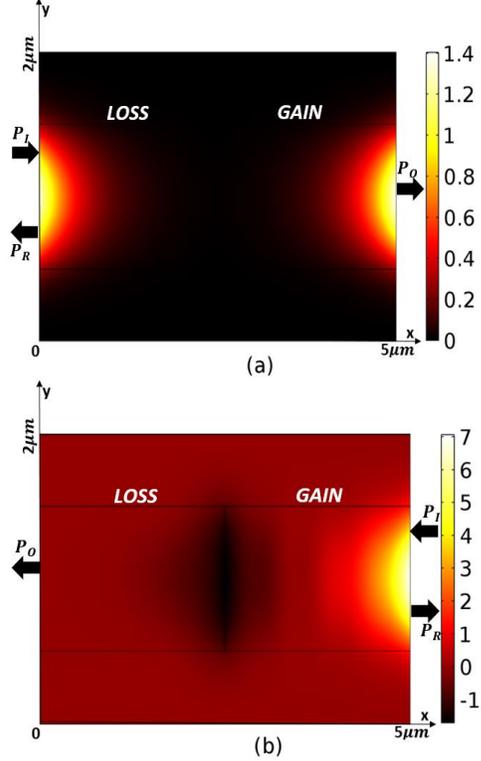

**Fig. 9.** Contour plot of 'Time-averaged power flow (in the unit of $10^6 W/m^2$)' for initial excitation of (a) loss port and (b) gain port when $n_I = 0.25$ (above PT threshold).

It should be noted that perfect transmission is really sensitive to the thickness or the optical constants to the layers. Next, we show how, with judicious choice of $n_I$ and the length of the active layers and the passive nano-film, it is possible to engineer the spectral singularity of the S-matrix. These spectral singularities enable amplified transmission at the desired wavelength regime [30]. For our calculations, we have assumed the following parameters: $a = 0.4$ μm , $b = 2.3$ μm and $\lambda = 1550$ nm . The structure is excited from the loss port only and the total length of the structure is left unchanged. Fig. 10(a) plots the eigen-spectra of the transfer matrix as a function of the loss/gain parameter $n_I$. We find that, in the region $n_I = 0.18 - 0.19$, the imaginary component of the eigenvalues flips its sign. We mean to say that the negative imaginary component of the eigenvalues becomes positive and vice-versa. At the same time, in Fig. 10(b), we can see that the transmission coefficient of the structure shows a sharp peak in the same region. The reason for this can be found in Fig. 11(a), wherein we see that the eigen-spectra of the S-matrix show a sharp dip. This is referred to as a spectral singularity of the S-matrix [30]. Now, if we choose $n_I \approx 0.18$, we obtain highly amplified transmitted radiation at 1550 nm, as illustrated in Fig. 11(b). The entire IR spectrum is blocked except the 1550 nm wavelength. Thus, we infer that the proposed structure could indeed be used as an IR filter.

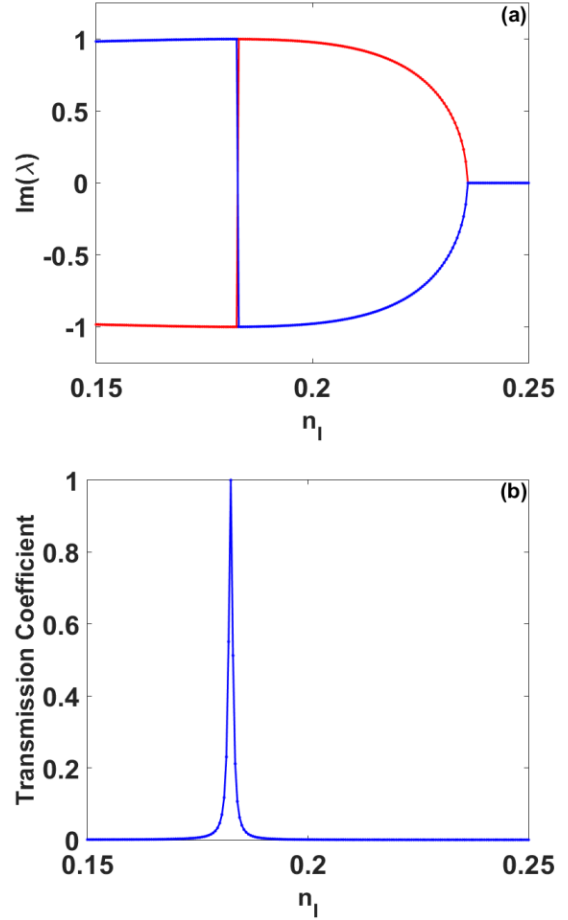

**Fig. 10.** (a) Eigen spectra of the transfer matrix ($\lambda = 1550$ nm) (b) Transmission Coefficient ($T'$) v/s Gain/Loss Coefficient $n_I$.

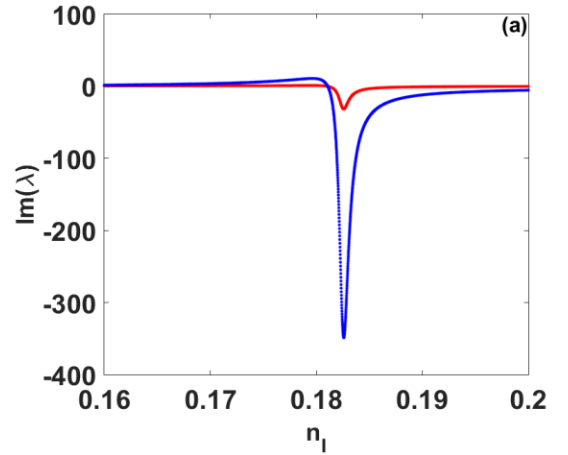

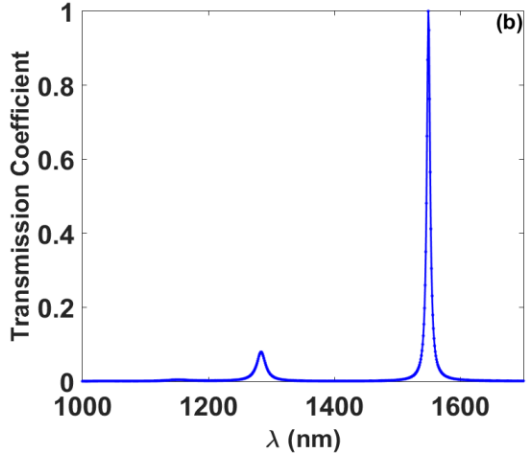

**Fig. 11.** (a) Eigen spectra of the S-matrix ($\lambda = 1550$ nm) (b) Transmission Coefficient ($T'$) v/s Wavelength ($n_I = 0.18$).

To confirm this inference further, we check the transmittance of the structure at the following wavelengths: red ($\lambda = 650$ nm), blue ($\lambda = 475$ nm) and green ($\lambda = 550$ nm) as a function of the gain/loss coefficient, $n_I$. We observe, as illustrated in Fig. 12, that the transmissivity vanishes for all these wavelengths as $n_I \to 0.18$. This clearly indicates that, only the wavelength, for which we observe spectral singularity in the S-matrix eigen-spectra, will be transmitted.

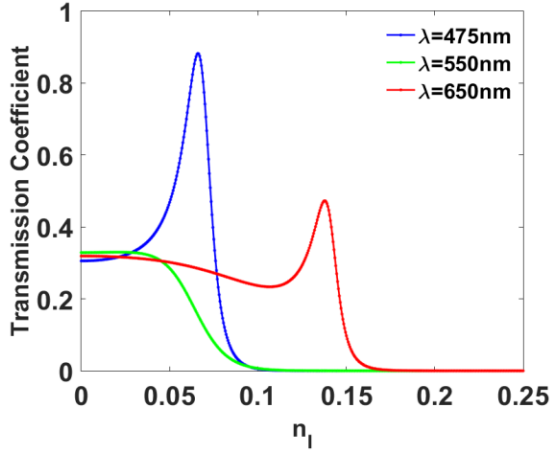

**Fig. 12.** Transmission coefficient ($T'$) v/s $n_I$ for red, blue and green wavelength.

Another interesting feature that we observed from our calculations is the existence of multiple $PT$ threshold. We have chosen the following parameter: $a = 0.6$ μm, $b = 2.2$ μm and $\lambda = 1550$ nm. For our choice of parameter values, we find (from Fig. 13(a)) that the eigen-spectra consist of three $PT$ thresholds, viz. $n_I \approx 0.247, 0.4095$ and $0.4280$. On top of that, we can see that the imaginary component of the eigenvalues flips its sign in the region $n_I = 0.41 - 0.43$. One important inference that we can draw from this is that the length of the nano-film and the silica layers plays a major role in the existence of the $PT$ threshold and also in the generation of spectral singularity of the S-matrix (as seen in Fig. 11(a)). To further validate our inference, we have shown in Fig. 13(b) the transmission coefficient v/s $n_I$ followed by the resonance spectrum in Fig. 14.

Our findings have confirmed the fact that our structure could indeed be utilized for perfect transmission of wavelength at $1550 nm$. With judicious choice of parameters and proper engineering of spectral singularities in the structure, it is possible to extend the proposed scheme for other wavelength regime as well.

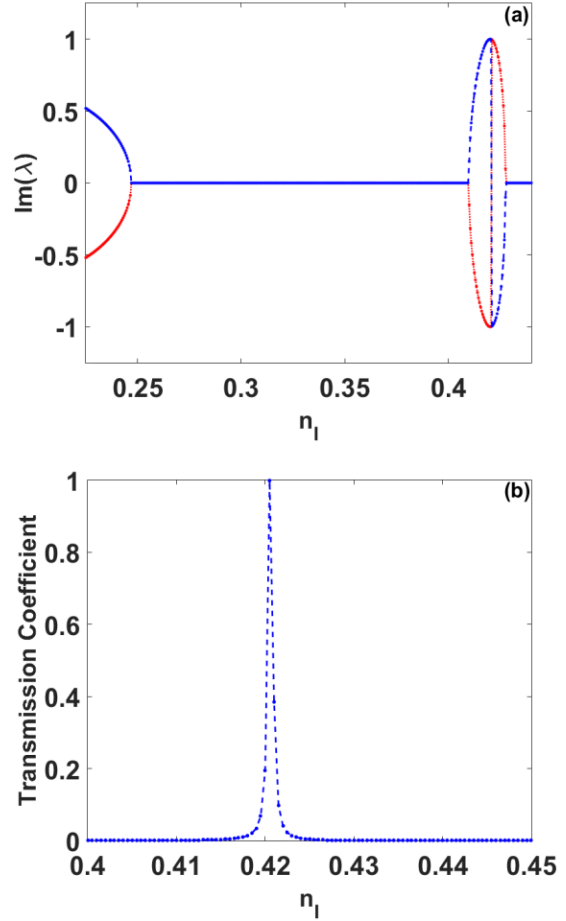

**Fig. 13.** (a) Eigen spectra of the transfer matrix ($\lambda = 1550$ nm) (b) Transmission Coefficient ($T'$) v/s Gain/Loss Coefficient $n_I$.

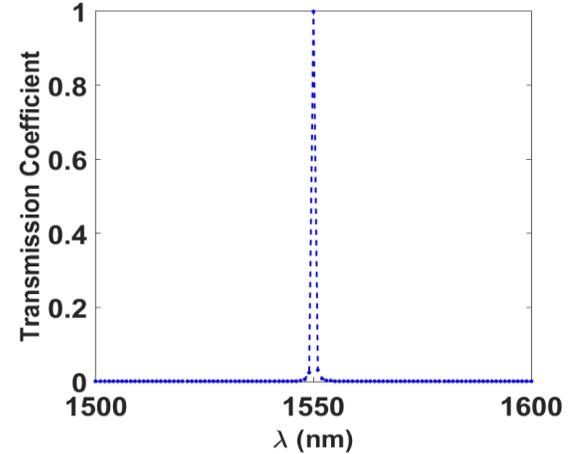

**Fig. 14.** Transmission Coefficient ($T'$) v/s Wavelength ($n_I = 0.42$).

## 4. CONCLUSION

In conclusion, we propose a parity time symmetric dielectric-nano-film-dielectric multilayered structure that could facilitate highly amplified transmission of optical power in the infrared spectrum. It is anticipated that, owing to the simplicity, the proposed structure and its variants could be practically implemented with ease. The analysis is carried out assuming the dielectric to be made of silica while the nano-film of sapphire. We have theoretically studied our model using the transfer matrix formalism. The theoretical results are validated by FDTD numerical simulation. Further, we show that, with judicious choice of the structure parameters, such as: length of the dielectric and the film layer and loss-gain parameters, it is possible to achieve highly amplified single wavelength transmission. The proposed structure may be useful for various applications in various light-based photonic and optoelectronics devices.

**Funding.** J.P.D. would like to thank MHRD, Govt. of India for financial support through a fellowship and A.K.S. would like to acknowledge the financial support from DST-SERB, Government of India (Grant No. SB/FTP/PS-047/2013).